\documentclass{ws-procs9x6}

\usepackage{graphicx,multirow,xspace,amsmath,amssymb}

\newcommand \sqsn{\mbox{$\sqrt{s_{_{NN}}}$}\xspace}

\def\mean#1{\ensuremath{\left<#1\right>}}

\newcommand{\sqrtsnn}{\ensuremath{\sqrt{s_{_{NN}}}}\xspace}
\newcommand{\UA}{\ensuremath{U_{A}(1)}\xspace}
\newcommand{\lambdas}{\ensuremath{\lambda_{*}}\xspace}
\newcommand{\lambdasmax}{\ensuremath{\lambda_{*}^{max}}\xspace}
\newcommand{\lamfrac}{\ensuremath{\lambdas(\mT)/\lambdasmax}\xspace}
\newcommand{\mT}{\ensuremath{m_\mathrm{T}}\xspace}
\newcommand{\pT}{\ensuremath{p_\mathrm{T}}\xspace}
\newcommand{\etap}{\ensuremath{{\eta^\prime}}\xspace}
\newcommand{\metap}{\ensuremath{m_{\etap}}\xspace}
\newcommand{\meps}{\ensuremath{m_{\etap}^{*}}\xspace}
\newcommand{\Tfo}{\ensuremath{T_{FO}}\xspace}
\newcommand{\Teff}{\ensuremath{T_{eff}}\xspace}
\newcommand{\Tcond}{\ensuremath{T_{cond}}\xspace}
\newcommand{\Binv}{\ensuremath{B^{-1}}\xspace}
\newcommand{\uT}{\ensuremath{\mean{u_\mathrm{T}}}\xspace}
\newcommand{\GeV}{\ensuremath{\mathrm{GeV}}\xspace}
\newcommand{\MeV}{\ensuremath{\mathrm{MeV}}\xspace}
\newcommand{\pip}{\ensuremath{\pi^+}\xspace}
\newcommand{\pim}{\ensuremath{\pi^-}\xspace}

\begin{document}

\title{$U_A(1)$ symmetry restoration \\
from an in-medium $\eta^\prime$ mass reduction \\
in $\sqrt{\lowercase{s}_{NN}} = 200$ GeV A\lowercase{u}+A\lowercase{u} collisions
\footnote{\uppercase{T}his work is supported by the 
\uppercase{H}ungarian \uppercase{OTKA} grant \uppercase{NK73143} and
by \uppercase{HAESF}, the \uppercase{H}ungarian-\uppercase{A}merican 
\uppercase{E}nterprise \uppercase{S}cholarship \uppercase{F}und.}}

\author{T. CS\"ORG\H{O} } 

\address{
Dept. Physics, Harvard University, \\
17 Oxford Street, Cambridge, MA 02138, USA, \\ 
and \\
MTA KFKI RMKI, \\
H-1525 Budapest 114, P.O.Box 49, Hungary \\ 
E-mail: csorgo@rmki.kfki.hu}

\author{R. V\'ERTESI and J. SZIKLAI}

\address{MTA KFKI RMKI, \\ 
H-1525 Budapest 114, P.O.Box 49, Hungary \\
E-mails: vertesi@rmki.kfki.hu, sziklai@rmki.kfki.hu}  

\maketitle

\abstracts{
A reduction of the mass of the $\eta^\prime$ (958) meson may signal 
restoration of the $U_A(1)$ symmetry in a hot and dense hadronic matter,
corresponding to the return of the 9th, "prodigal" Goldstone boson.  
We report on an analysis of a combined PHENIX and STAR data set 
on the intercept parameter of the two-pion Bose-Einstein correlation functions, 
as measuremed in \sqrtsnn = 200 GeV Au+Au collisions at RHIC. 
To describe this combined PHENIX and STAR dataset, 
an in-medium $\etap$ mass reduction of at least 200 MeV is needed,
at the 99.9 \% confidence level in a broad model class of resonance abundances. 
}

\section{Introduction}

Although the quark model exhibits a $U(3)$ chiral symmetry in the limit of massless up, down and strange quarks,
and in principle 9 massless Goldstone modes are expected to appear when this symmetry is broken, 
only 8 light pseudoscalar mesons are observed experimentally. This puzzling mystery is resolved by 
the Adler-Bell-Jackiw \UA anomaly: instantons tunneling between topologically different QCD vacuum 
states explicitely break the \UA part of the $U(3)$ symmetry. 
Thus the 9th Goldstone boson is expected to be massive, and is associated with 
the \etap meson, which has a mass of 958 MeV, approximately twice that of the other pseudoscalar mesons.

In high energy heavy ion collisions at RHIC, a hot and dense medium is created. 
Recent measurements of the direct photon spectrum in \sqrtsnn = 200 GeV Au+Au collisions
indicate~\cite{PHENIX-directphotons}, that the initial temperature in these reactions is at least 300 MeV, 
while hadrons as we know them may not exist above the Hagedorn temperature 
of $T_H \approx 170$ MeV~\cite{Hagedorn}.  Thus the matter created
in heavy ion collisions at RHIC is hot enough to be a quark-gluon plasma~\cite{PHENIX-directphotons}. 
Detailed analysis of the properties of this matter indicate that it flows like  a perfect fluid~\cite{PHENIX-WhitePaper},
and scaling properties of the elliptic flow indicate scaling with the number of constituent quarks~\cite{PHENIX-v2scaling},
hence this matter is sometimes referred to as a strongly interacting Quark-Gluon Plasma (sQGP)~\cite{shuryak-sQGP},
or, in more direct terms, a perfect fluid of quarks~\cite{PHENIX-WhitePaper}.
          
After this perfect fluid of quarks rehadronizes, a hot and dense hadronic matter may be created, where the \UA
symmetry of the strong interactions  may temporarily be restored \cite{kunihiro,kapusta,huang}.
Recent lattice QCD calculations indicate that such chirally symmetric but hadronic matter may exist
below the critical temperature for quark deconfinement~\cite{Fodor:2009ax}.
In such a medium, the mass of the $\etap(958)$ mesons may be reduced to its  quark model value of about 500 MeV, 
corresponding to the return of the ``prodigal" 9th Goldstone boson~\cite{kapusta}. 
Here we report on an indirect observation of such an in-medium $\etap$ mass modification based 
on a detailed analysis of PHENIX and STAR charged pion Bose-Einstein correlation (BEC) data~\cite{phnxpub,starpub}.

The abundance of the $\etap$ mesons with reduced mass may be increased at low \pT, 
by more than a factor of 10. One should emphasize that the \etap (and $\eta$) mesons almost 
always decay after the surrounding hadronic matter has frozen out, due to their small 
annihilation and scattering cross sections, and their decay times that are much longer than the
characteristic 5-10 fm/c decoupling times of the fireball created in high energy heavy ion 
collisions. Therefore one cannot expect a direct observation of the mass shift of the $\etap$ 
(or $\eta$) mesons: all detection possibilities of their in-medium  mass  modification have 
to rely on their enhanced production.

An enhancement of low transverse momentum \etap mesons contributes to an enhanced production of soft charged pions mainly through the $\etap \rightarrow \eta + \pi^+ + \pi^- \rightarrow (\pi^+ + \pi^0 + \pim) + \pip + \pim$ decay chain and also through other, less prominent channels. As the \etap decays far away from the fireball, the enhanced production of pions in the corresponding halo region will reduce the strength of the Bose-Einstein correlation between soft charged pions.
The transverse mass ($\mT = \sqrt{m^2 + \pT^2}$) dependence of the extrapolated intercept parameter 
\lambdas of the charged pion Bose-Einstein correlations 
was shown to be an observable that is sensitive to such an enhanced $\etap$ multiplicity,
as pointed our first in Ref.~\cite{vance} and discussed in Ref.~\cite{Csorgo:1999sj}.  
The predicted decrease of $\lambdas(\mT)$ data at low transverse mass 
has  been observed both by PHENIX~\cite{phnxpub} and STAR~\cite{starpub,Abelev:2009tp} at RHIC.

\begin{figure}[htb]
\centerline{\epsfxsize=12cm\epsfbox{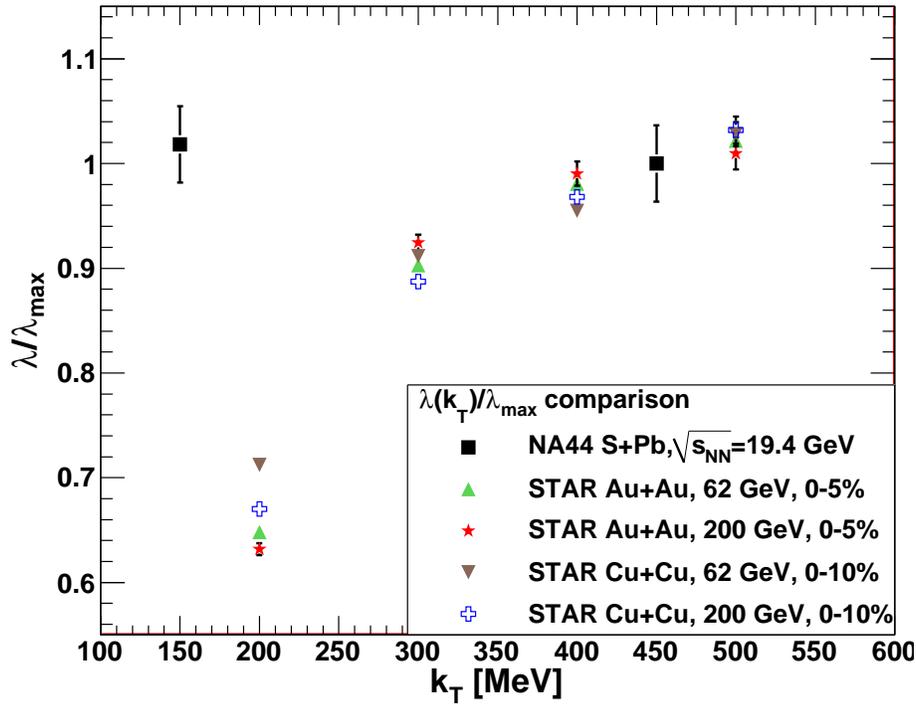}}   
\caption{
The energy and system size dependence of the relative intercept parameter in the 
NA44 S+Pb and STAR Cu+Cu and Au+Au datasets.
\label{f:lambdarel-excite}
}
\end{figure}

Figure \ref{f:lambdarel-excite} indicates that a ``hole" at low transverse
mass, in the region characteristic to the
 $\eta^\prime \rightarrow \eta + \pi^+ + \pi^- 
\rightarrow  \eta^\prime \rightarrow (\pi^+ \pi^0 \pi^- )+ \pi^+ + \pi^- $ decay chain 
is not present in the S+Pb data set at CERN SPS at \sqrtsnn = 19.4 GeV energy~\cite{Beker:1994qv}, 
however, it is present  and becomes slightly deeper as the system size and the colliding energy is increased
from Cu+Cu to Au+Au and from \sqrtsnn = 62.4 GeV to 200 GeV~\cite{Abelev:2009tp}.

\section{Modeling and analysis method}
Our main analysis tool was a Monte-Carlo simulation of the transverse mass dependence of
the long lived resonance multiplicities including the possibility of an enhanced $\etap$
production at low transverse momentum, due to a partial in-medium \UA restoration and a related $\etap$ mass modification. 
This model and the related reduction of the effective intercept parameter of the two-pion Bose-Einstein
correlation function was proposed first in ref.~\cite{vance} and detailed recently in Refs.~\cite{csvsz-PRL,Vertesi:2009ca,Vertesi:2009wf}.

In thermal models, the production cross sections of the light mesons are exponentially suppressed by the mass. 
Hence one expects about two orders of magnitude less \etap mesons from the freeze-out than pions. 
This suppression, however, may be moderated as a consequence of a possible \etap mass reduction, 
and the $\etap$ mesons may show up in an enhanced number. 
The number of in-medium $\etap$ mesons is calculated with an improved Hagedorn formula
yielding the following \etap enhancement factor:
\begin{equation}\label{eq:prietamtdist}
f_\etap=\left(\frac{\meps}{\metap}\right)^\alpha e^{- \frac{\metap-\meps}{\Tcond}}.
\end{equation}
This formula includes a prefactor with an expansion dynamics dependent exponent $\alpha\approx 1-d/2$ for an expansion in $d$ effective dimensions~\cite{Csorgo:1995bi}. As a default value, $\alpha = 0$ was taken~\cite{vance} and, for the systematic investigations,
this parameter was varied between $ -0.5 \le \alpha \le 0.5$.
Other model parameters and their investigated ranges  are described as follows:
$\Tcond$ in the above formula corresponds to 
the temperature of the medium when the in-medium modified $\etap$ mesons are formed; its default value
was taken to be $\Tcond = 177$ MeV~\cite{vance} and varied systematically
between 140 and 220 MeV. 
Resonances with different masses were simulated with a mass dependent slope parameter
$\Teff = \Tfo + m \langle u_T\rangle^2 $, where the default values of $\Tfo = 177$ MeV
and $\langle u_T\rangle = 0.48$~\cite{Adler:2003cb} were utilized and 
systematically varied in the range of 100 MeV  $\le \Tfo \le $ 177 MeV and 0.40  $\le \langle u_T\rangle \le $ 0.60~.

Once produced, the \etap is expected to be decoupled from other hadronic matter, since its annihilation and scattering cross sections are very small~\cite{kapusta}.
If the \etap mass is reduced in the medium, the observed \etap spectrum will consist of two components.
If the \pT of the \etap is large enough, it can get on-shell and escape. 
This will produce a thermal component of the spectrum. 
Energy conservation at mid-rapidity implies ${m_{\etap}^*}^2+{p_{T,\etap}^{*}}^2={m_{\etap}}^2+{p_{T,\etap}}^2$.
(In the latter equation the quantities marked with an asterisk denote the properties of the in-medium \etap, while the ones without an asterisk refer to the free \etap.)
On the other hand, \etap-s with ${p_{T,\etap}^{*}} \le \sqrt{ {m_{\etap}^*}^2-{m_{\etap}}^2 }$ will not be able to leave the hot and dense region through thermal fluctuation since they cannot compensate for the missing mass~\cite{kapusta,huang}, and thus will be trapped in the hot and dense region until it disappears. As the energy density of the medium is dissolved, the effect of QCD instantons increases and the trapped \etap mesons regain their free mass and appear at low \pT. 

In our recent works of Refs.~\cite{csvsz-PRL,Vertesi:2009ca,Vertesi:2009wf}, 
we improved on earlier simulations of Ref.~\cite{vance}, that considered the 
trapped \etap mesons to leave the dissolving medium with a negligible \pT . That earlier approach 
resulted in a steep hole in the extrapolated intercept parameter 
$\lambdas(\mT)$ at a characteristic transverse mass of $\mT\le 250\ \MeV$~\cite{vance,Csorgo:1999sj,phnxpre}. 
In that simplified scenario the only free parameter was the in-medium \etap mass, 
determining the depth of the observed hole. 
In a recent analysis, summarized here,  the \etap-s from the decaying condensate were given a 
random transverse momentum, following Maxwell-Boltzmann statistics with an 
inverse slope parameter \Binv{}, which was necessary to obtain a quality description of the 
width and the slope of the $\lambdas(\mT)$ data of PHENIX and STAR in the $\mT \approx 300$ MeV region.
Physically, $\Binv$ is limited by $\Tfo$, so the trapped \etap -s may gain only moderate transverse momenta.
Hence, the enhancement mostly appears at low $p_T$~\cite{kunihiro,kapusta,huang} just as in the first simulations.
However, now the slope of ``hole" of the $\lambdas(\mT)$ curve is determined by $\Binv$,
 and, for certain values of the model parameters, the data can be reproduced quantitatively.
(The \lambdas values, actually used in the presented analysis, and their total errors are discussed in details in Ref.~\cite{Vertesi:2009wf}. Here $\lambdas^\mathrm{max}$ is the $\lambdas(\mT)$ value taken at $\mT=0.7\ \GeV$, with the exception of the STAR data, where the data point at the highest $\mT=0.55\ \GeV$ is considered. Note that the \mT dependency of the  $\lambdas(\mT)$ measurements in the 0.5-0.7 GeV region is very weak.)

\begin{figure}[htb]
\centerline{\epsfxsize=12cm\epsfbox{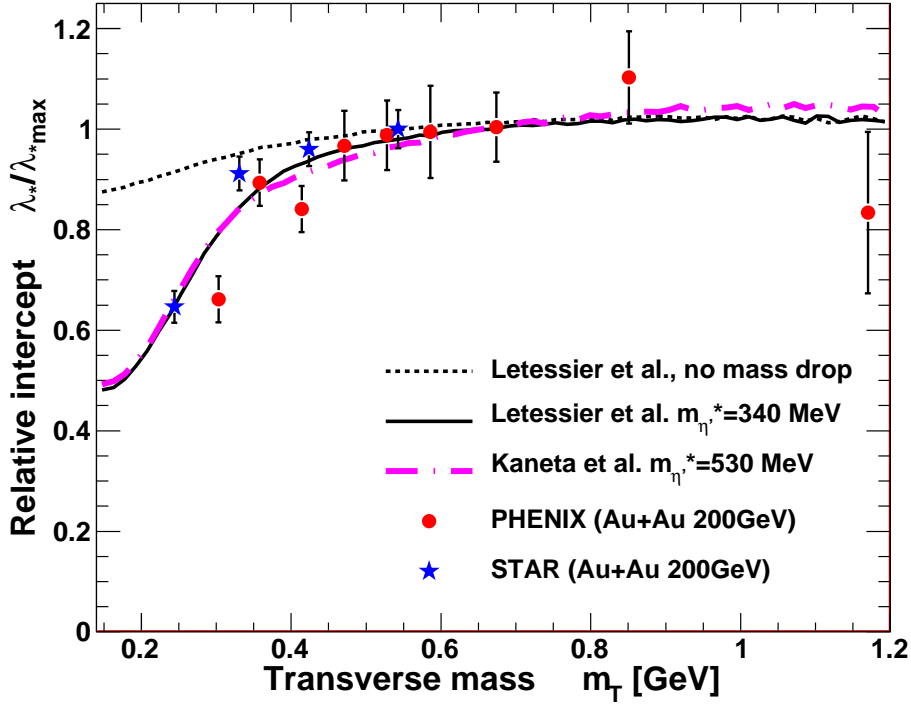}}   
\caption{
The transverse mass dependence of the relative intercept parameter in the 
PHENIX and STAR dataset is reproduced with an in-medium mass modification of the
$\eta^\prime$ mesons using two different resonance models as input.
The same resonance models, but without in-medium mass modification, cannot explain these
datasets.
\label{f:lambdarel}
}
\end{figure}

We have investigated a broad class of models of resonance production, 
including two different models that produce resonances without assuming local thermalization: 
FRITIOF~\cite{fritiof} and UrQMD~\cite{urqmd}. 
Resonance decays, including decay chains, were simulated with JETSET 7.4~\cite{Sjostrand:1995iq}.

The FRITIOF~\cite{fritiof} Monte Carlo model, based on superposition of nucleon-nucleon collisions and the Lund string fragmentation model, cannot describe the behavior seen in \lamfrac even when an arbitrary \etap mass modification is considered. On the other hand, hadronic cascade based UrQMD~\cite{urqmd}, as well as the quark coalescence model ALCOR~\cite{alcor} and the thermal resonance production models of refs.~\cite{kaneta,stachel,rafelski}, provide a successful fit in a certain range of the in-medium \etap masses.
The main difference between the thermal models that we utilized was in those resonance multiplicities that are not yet measured well: ref.~\cite{kaneta} predicts a factor of 1.6 more $\eta$-s and a factor of 3 more $\etap$-s than the models of ref.~\cite{rafelski,stachel}. The relevant resonance fractions of these models are detailed in Table V of ref.~\cite{Vertesi:2009wf}.

The dotted line in Fig.~\ref{f:lambdarel} indicates a scenario without an in-medium \etap mass reduction,
 while the dot-dashed and solid lines show the enhancement required to describe the dip in the low \mT region of \lambdas corresponding to the resonance multiplicities of Refs.~\cite{kaneta,rafelski}, respectively.

\begin{figure}[htb]
\centerline{\epsfxsize=12cm\epsfbox{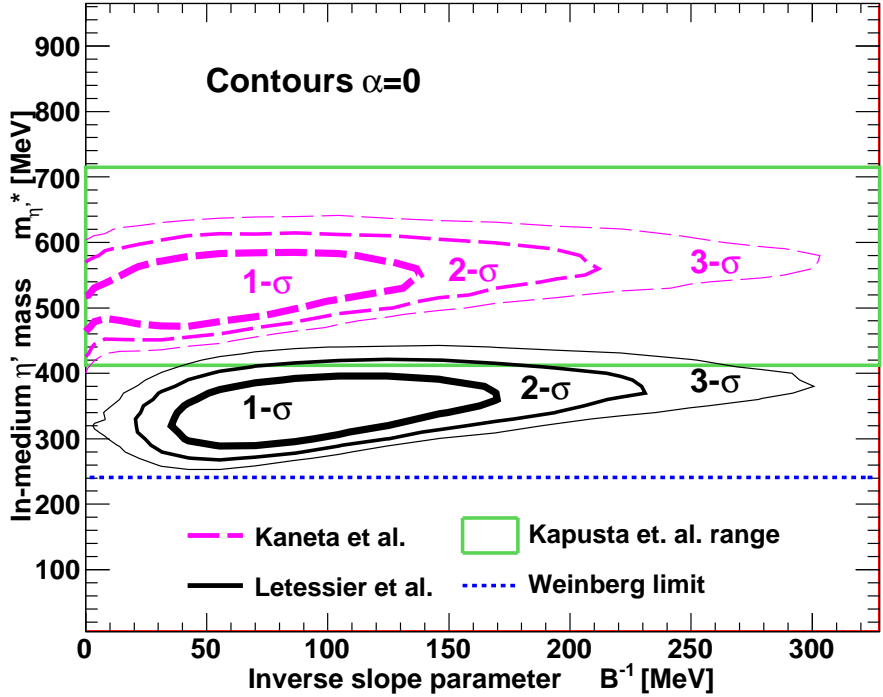}}   
\caption{
Standard deviation contours on the (\Binv{}, \meps) plain, obtained from 
\lamfrac\ of Monte Carlo simulations based on particle multiplicities
using two different models for hadronic resonances. 
The region between the horizontal solid lines indicates the theoretical range 
predicted by quark model considerations, while  
the dotted horizontal line stands for Weinberg's lower limit.
\label{f:xcont}
}
\end{figure}

Based on extensive Monte-Carlo simulations, $\chi^2$ of the fits to the data of Fig.~\ref{f:lambdarel}
 was computed as a function of $m_{\etap}^*$ and $B^{-1}$ for each resonance model and each fixed 
value of model parameters of $\alpha$, $\Tcond$, $\Tfo$ and $\uT$. 
The best values for the in-medium mass of $\eta^\prime$ mesons are in, or slightly below, the range 
$\sqrt{\frac{1}{3}(2 m_{\rm K}^2 + m_\pi^2)}\le\meps\le \sqrt{2 m_{\rm K}^2 - m_\pi^2}$
predicted in  ref.~\cite{kapusta}, while all are above the lower limit of 
$\meps\ge\sqrt{3}m_\pi$ 
given by ref.~\cite{Weinberg:1975ui}.
The \lamfrac simulations for the best fits of two characteristic models are compared to the no-mass-drop scenario on Fig.~\ref{f:lambdarel}, while the 1, 2 and 3-$\sigma$ parameter boundaries are indicated in Fig.~\ref{f:xcont}. 
Those models that describe both PHENIX and STAR \lamfrac data in a statistically 
acceptable manner with the assumption of a sufficiently large in-medium $\eta^\prime$ mass reduction 
are all used for the estimation of systematics. The key parameters of the best fits are listed in Table~\ref{tab:modelsum}.

\section{Results }
We have used different input models and setups to map the parameter space for a twofold goal:
to determine, at least how big $\eta^\prime$ in-medium mass reduction is needed to be able to describe these datasets,
and also to determine, what are the best values of the in-medium mass modification of the $\eta^\prime$ mesons. 
Utilizing our indirect method, we have also reconstracted the 
transverse mass dependent spectrum of these $\eta^\prime$ mesons. 

\subsection{Lower limit on the in-medium $\eta^\prime$ mass reduction }
 We excluded certain regions where a statistically acceptable fit to the data is not achievable, thus we can give a lower limit on the \etap mass modification. 
At the 99.9 \% confidence level, corresponding to a more than 5-$\sigma$ effect, at least 200 MeV in-medium decrease of the mass of the $\eta^\prime(958)$ meson was needed to describe both STAR 0-5 \% central and PHENIX 0-30\% central Au+Au  data on $\lamfrac$ in $\sqrt{s_{NN}} = 200$ GeV Au+Au collisions at RHIC, in the considered model class.

\subsection{Best value of the in-medium $\eta^\prime$ mass reduction }
We have determined the best values and errors of the fitted \meps and \Binv parameters. 
The best simultaneous description of PHENIX~\cite{phnxpub} and STAR~\cite{starpub} 
relative intercept parameter data is achieved with an $\etap$ mass that is dramatically 
reduced in the medium created in central Au+Au collisions at RHIC from its vacuum value of 958 MeV to
$340{+50\atop -60}{+280\atop -140}\pm{45}$ MeV.  The first error here is the statistical one determined by 
the 1-$\sigma$ boundaries of the fit. The second error is from the choice of the resonance model and the parameters ($\alpha$, $\Tcond$, $\Tfo$ and $\uT$) of the simulation. The third error is the systematics resulting from slightly different PHENIX and STAR centrality ranges, particle identification and acceptance cuts. These effects have been estimated with Monte-Carlo simulations, detailed in ref.~\cite{Vertesi:2009wf}, not to exceed 9.8\%, 7\% and 3\% respectively.
The main source of systematic errors is the choice of the resonance models. This is due to 
the unknown initial $\etap$ multiplicity, hence models like ref.~\cite{kaneta} with larger initial \etap abundances require smaller in-medium \etap mass modification, as compared to the models of ref.~\cite{stachel,rafelski}.

\subsection{Transverse mass spectrum of $\eta^\prime$ mesons}
\begin{figure}[htb]
\centerline{\epsfxsize=12cm\epsfbox{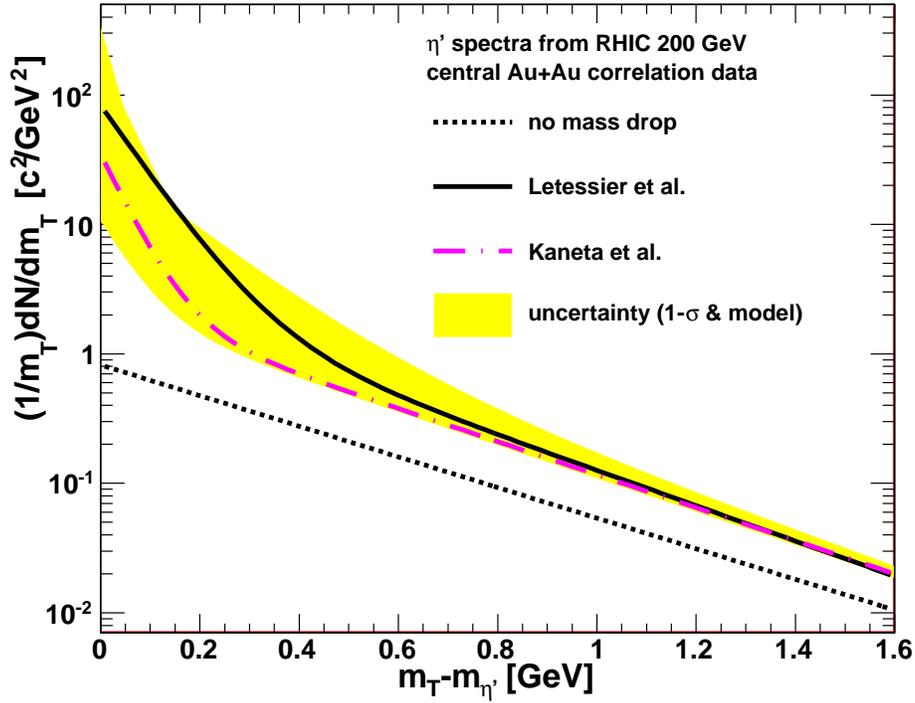}}   
\caption{
The transverse mass dependent spectrum  of the $\eta^\prime$ mesons,
obtained using two different resonance models as input.
The band indicates the systematic error, obtained from varying the 
resonance models as discussed in the text.
\label{f:etap-spectrum}
}
\end{figure}

In addition to the characterization of the in-medium $\etap$ mass modification, 
the transverse momentum spectra of the $\eta$ and $\etap$ mesons have also been reported in
Ref.~\cite{csvsz-PRL}.  Fig. \ref{f:etap-spectrum} indicates the reconstructed spectrum of \etap mesons
in \sqrtsnn = 200 GeV Au+Au collisions, for simulations based on 
resonance abundances of Refs.~\cite{rafelski,kaneta}. 
Normalization was carried out with respect to the 
$\eta^\prime$ multiplicity of the model described  in Ref.~\cite{kaneta}.
The spectrum of Fig. ~\ref{f:etap-spectrum} features a characteristic low transverse momentum enhancement. 
Although PHENIX measured before the $\eta$ spectrum in the $\pT\ge 2\ \GeV$ region~\cite{Adler:2006bv},
as far as we know the spectrum of the $\etap$ particles has not been 
determined before in \sqrtsnn = 200 GeV Au+Au collisions at RHIC.  

\begin{figure}[htb]
\centerline{\epsfxsize=12cm\epsfbox{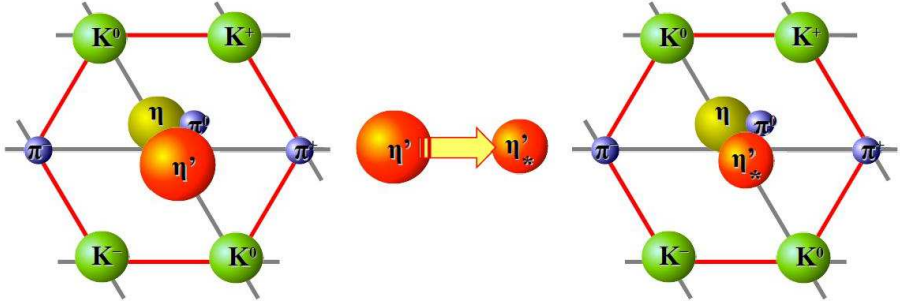}}   
\caption{
The left hand side plot indicates the 9 pseudoscalar mesons,
the size of the peebles is proportional to the mass, respectively.
The central plot indicates that the $\eta^\prime$ 
meson reduces its mass in a hot and dense hadronic medium 
created in Au+Au collisions at RHIC.
The right-hand side plot indicates that after the $U_A(1)$ symmetry
is effectively restored, the mass of the  $\eta$ and the $\eta^\prime$  
mesons will be similar. 
\label{f:etaprime-ua1}
}
\end{figure}

The restoration of the $U_A(1)$ symmetry and the symmetry between mass of the $\eta$ and the  
$\eta^\prime$ mesons is illustrated on Fig.~\ref{f:etaprime-ua1}. 

\begin{table}[htbp]
\tbl{
Best fits of $\meps$ and $\Binv$ for different resonance multiplicity models, 
followed by $\chi^2/\mbox{\it NDF}$ 
and the corresponding confidence level (CL).
The integrated  $\etap$ and $\eta$ enhancement factors $f_\etap$ and $f_\eta$ are 
followed by the 5-$\sigma$ limits of maximum in-medium masses.
Errors on \meps values represent 1-$\sigma$ boundaries,
while the 5-$\sigma$ limits include systematic errors too.
}
{\begin{tabular}{c c c c c c c}
\hline \\[-1.7ex]
Resonance & \meps  & \Binv  & \multirow{2}{*}{ 
  $\frac{\displaystyle\strut \chi^2}{\displaystyle\strut \mbox{\it NDF}}$ 
(CL \%)} & $f_\etap$ & $f_\eta$ & 5-$\sigma$ limit \\
  model   &  (MeV) & (MeV)  &  &  & & $\meps$ (MeV) \\[0.6ex]
\hline \\[-1.7ex]
ALCOR~\cite{alcor}
& $490{+60\atop-50}$ & 42 & 20.2/11 (4.29) & 43.4 & 5.25 & $\le$ 700 \\
Kaneta~\cite{kaneta}
& $530{+50\atop-50}$ & 55 & 22.8/11 (4.12) & 25.6 & 3.48 & $\le$ 730 \\
Letessier~\cite{rafelski}
& $340{+50\atop-60}$ & 86 & 18.9/11 (6.35) & 67.6 & 4.75 & $\le$ 570 \\
Stachel~\cite{stachel}
& $340{+50\atop-60}$ & 86 & 18.9/11 (6.38) & 67.6 & 4.97 & $\le$ 570 \\
UrQMD~\cite{urqmd}
& $400{+50\atop-40}$ & 86 & 18.9/11 (6.14) & 45.0 & 7.49 & $\le$ 660 \\[1.0ex]
\hline
\end{tabular}
\label{tab:modelsum}
}
\end{table}

\section{Discussion}
Detailed analysis of the STAR and PHENIX \lamfrac dataset recorded at 7.7, 9.2, 11.5, 39 and 62.4 GeV during 2010 has just been started~\cite{Abelev:2009bw}, marking the beginning of the RHIC energy scan program.

At present, detailed data are available from the NA44 collaboration at $\sqsn=19.4\ \GeV$~\cite{Beker:1994qv} 
as well as from the STAR collaboration at $\sqsn=62.4$ and 200 \GeV Cu+Cu and Au+Au collisions, the latter at different centrality classes within the 0\%--80\% range~\cite{Abelev:2009tp}. 
The NA44 data at $\sqsn=19.4\ \GeV$ does not feature an \etap mass drop effect. A positive sign of the \etap mass modification is apparent in each case of the STAR datasets, indicating that the mass modification effect is nearly at maximum in $\sqsn=200\ \GeV$ Au+Au collisions and reduces with decreasing centrality, colliding energy and system size. We have estimated the magnitude of the system size and energy dependence between 62.4 GeV Cu+Cu and 200 GeV Au+Au collisions to be not larger than 15\%, which is substantially less than the dominant systematic error coming from the choice of the resonance model.

The {\it dilepton spectrum} has been measured recently in minimum bias Au+Au collisions at \sqsn = 200 GeV,
and a large enhancement was observed in the low invariant mass region $m_{\rm ee}<1\ \GeV$~\cite{Adare:2009qk}.
Low transverse mass enhancement of the \etap and $ \eta$ production results in dilepton enhancement just in this kinematic range~\cite{kapusta}. 
Estimations using the enhancement factors in Table~\ref{tab:modelsum} indicate that the observed in-medium \etap mass drop is indeed a promising candidate to explain this dilepton excess.

PHENIX recently reported a two-component transverse momentum spectrum in dilepton channel {\it direct photon measurements}~\cite{Adare:2009qk}, which provides an additional testing possibility to constrain the two component structure of the \etap spectra reported here.

\section{Summary}
Our report presents a statistically significant, 
indirect observation of an in-medium mass modification 
of the $\eta^\prime$ mesons in $\sqsn=200\ \GeV$ Au+Au collisions at RHIC.
These results were recently published in Refs.~\cite{csvsz-PRL,Vertesi:2009ca,Vertesi:2009wf}.
A similar search for in-medium $\etap$ mass modification provided negative result in S+Pb reactions at
CERN SPS energies~\cite{vance}.  
More detailed studies of the excitation function, 
the centrality and system size dependence of the $\lamfrac$ could provide important additional 
information about the onset and saturation of the partial \UA symmetry restoration in hot and dense hadronic matter. 
Studies of the low-mass dilepton spectrum and measurements of other decay channels 
of the $\etap$ meson may shed more light on the reported magnitude of the low $\pT$ $\etap$ enhancement 
and the related \UA symmetry restoration in high energy heavy ion collisions.

\section*{Acknowledgments}
We thank the Organizers of the Gribov 80 Memorial Workshop for creating an inspiring scientific atmosphere 
and providing an excellent setting for scientific discussions. We also would like to  thank to professors R.~J.~Glauber 
and Gy.~Wolf for inspiring and clarifying discussions. T.~Cs. is grateful to R.~J.~Glauber for his kind hospitality at the Harvard University. Our research was supported by Hungarian OTKA grant NK 73143. T. ~Cs. has also been supported by
a Senior Leader and Scholar Fellowship by the Hungarian American Enterprise Scholarship Fund (HAESF).

\end{document}